\iffalse Before sending further, pass this check list:
  [ ] find and resolve all "??" and "\ifnote"
  [ ] Spell-check
  [ ] order of using/defining abbreviations
  [ ] citation order (use RefTest)
  [ ] figure order & reference order
  [ ] Check LaTeX  output files (*.log) for warnings
  [ ] Check BibTeX output (screen) for warnings
  [ ] update PACS
  [ ] decide on color figures
  [ ] Re-read paper in the morning

After completing this list, if you made at least one correction,
re-do this check-list from the beginning, until no corrections will
be done.\fi

\documentclass[amsmath,amssymb,prb,preprintnumbers,showpacs,superscriptaddress,twocolumn]{revtex4}

\usepackage{graphicx}
\usepackage{dcolumn}          
\usepackage{bm}               
\usepackage[usenames]{color}  
\usepackage{ulem}             
\usepackage[colorlinks=true, linkcolor=black, citecolor=black, urlcolor=black]{hyperref} 
\usepackage{cancel}	      

\newcommand{\dxy}{$d_{x^2-y^2}$}

\newif\ifcom
\newif\ifdel

\begin{document}


\title{Phase-sensitive evidence for $d_{x^2-y^2}$ - pairing symmetry in the\\ parent-structure high-$T_c$ cuprate superconductor Sr$_{1-x}$La$_{x}$CuO$_{2}$}

\author{J.~Tomaschko}
\affiliation{%
  Physikalisches Institut -- Experimentalphysik II and Center for Collective Quantum Phenomena in LISA$^+$,
  Universit\"{a}t T\"{u}bingen,
  Auf der Morgenstelle 14,
  72076 T\"{u}bingen, Germany
}
\author{S.~Scharinger}
\affiliation{%
  Physikalisches Institut -- Experimentalphysik II and Center for Collective Quantum Phenomena in LISA$^+$,
  Universit\"{a}t T\"{u}bingen,
  Auf der Morgenstelle 14,
  72076 T\"{u}bingen, Germany
}
\author{V.~Leca}
\affiliation{%
  Physikalisches Institut -- Experimentalphysik II and Center for Collective Quantum Phenomena in LISA$^+$,
  Universit\"{a}t T\"{u}bingen,
  Auf der Morgenstelle 14,
  72076 T\"{u}bingen, Germany
}
\affiliation{%
  National Institute for Research and Development in Microtechnologies, 
  Molecular Nanotechnology Laboratory, 
  Erou Iancu Nicolae Str. 126A, RO-077190, Bucharest, Romania
}
\author{J.~Nagel}
\affiliation{%
  Physikalisches Institut -- Experimentalphysik II and Center for Collective Quantum Phenomena in LISA$^+$,
  Universit\"{a}t T\"{u}bingen,
  Auf der Morgenstelle 14,
  72076 T\"{u}bingen, Germany
}
\author{M.~Kemmler}
\affiliation{%
  Physikalisches Institut -- Experimentalphysik II and Center for Collective Quantum Phenomena in LISA$^+$,
  Universit\"{a}t T\"{u}bingen,
  Auf der Morgenstelle 14,
  72076 T\"{u}bingen, Germany
}
\author{T.~Selistrovski}
\affiliation{%
  Physikalisches Institut -- Experimentalphysik II and Center for Collective Quantum Phenomena in LISA$^+$,
  Universit\"{a}t T\"{u}bingen,
  Auf der Morgenstelle 14,
  72076 T\"{u}bingen, Germany
}

\author{D.~Koelle}
\author{R.~Kleiner}
\email{kleiner@uni-tuebingen.de}
\affiliation{%
  Physikalisches Institut -- Experimentalphysik II and Center for Collective Quantum Phenomena in LISA$^+$,
  Universit\"{a}t T\"{u}bingen,
  Auf der Morgenstelle 14,
  72076 T\"{u}bingen, Germany
}
%
\date{\today}

\begin{abstract}
Even after 25 years of research the pairing mechanism and -- at least for electron doped compounds -- also the order parameter symmetry of the high transition temperature (high-$T_c$) cuprate superconductors is still under debate. One of the reasons is the complex crystal structure of most of these materials. An exception are the infinite layer (IL) compounds consisting essentially of CuO$_2$ planes. Unfortunately, these materials are difficult to grow and, thus, there are only few experimental investigations. Recently, we succeeded in depositing high quality films 
of the electron doped IL compound Sr$_{1-x}$La$_{x}$CuO$_{2}$ (SLCO), with $x \approx 0.15$, and on the fabrication of well-defined grain boundary Josephson junctions (GBJs) based on such SLCO films.  Here we report on a phase sensitive study of the superconducting order parameter based on GBJ SQUIDs from a SLCO film grown on a tetracrystal substrate. Our results show that also the parent structure of the high-$T_c$ cuprates has \dxy-wave symmetry, which thus seems to be inherent to cuprate superconductivity.
\end{abstract}


\maketitle


Since the discovery of high transition temperature (high-$T_c$) superconductivity in cuprates\cite{Bednorz86}, tremendous work has been performed on these materials.
Researchers succeeded in increasing $T_c$ from initially 30\,K to 135\,K\cite{Kim95a,Kim95,Puzniak95} by synthesizing increasingly complex compounds. 
However, the microscopic mechanism causing high-$T_c$ superconductivity still has not been identified and is one of the biggest issues in solid state physics.
All these materials have in common that superconductivity resides in the copper oxide (CuO$_2$) planes where superconducting charge carriers, Cooper pairs, form.
An ``infinite layer'' (IL) cuprate consisting essentially of CuO$_2$ planes is therefore of fundamental interest for all questions addressing the basics of high-$T_c$ superconductivity.
In 1988, Siegrist \textit{et al.} succeeded in synthesizing such a simple cuprate, which is known as the ``parent structure'' of cuprate superconductors\cite{Siegrist88}.
Its CuO$_2$ planes are only separated by a single alkaline earth metal plane ($A$ = Ca, Sr or Ba), forming a $A$CuO$_2$ crystal.
Upon electron-doping, it turned out to be superconducting with maximum $T_c = 43\,$K\cite{Smith91,Er92,Jorgensen93,Ikeda93}.
%


A striking and highly debated feature of the cuprate superconductors is their unconventional order parameter symmetry.
Whereas for hole-doped cuprates \dxy-wave pairing has been established\cite{Scalapino95,vanHarlingen95,Tsuei00}, for electron-doped cuprates \cite{Armitage10} 
the issue is not yet completely settled.
After controversial discussion, the electron-doped $T'$-compounds $L_{2-x}$Ce$_x$CuO$_4$ ($L$ = La, Pr, Nd, Eu or Sm) have been shown to be predominant \dxy-wave superconductors by a number of phase-sensitive experiments\cite{Tsuei00a,Chesca03,Ariando05,Guerlich09}.
By contrast, for the parent compounds the pairing symmetry is essentially still unknown, since a variety of experimental tests yielded conflicting results\cite{Imai95,Chen02,Williams02,Zapf05,Liu05,Satoh08,Khasanov08,White08,Teague09,Fruchter10,Armitage10}.
Phase-sensitive tests, such as experiments on corner junctions\cite{Wollman93}, tricrystal rings\cite{Tsuei94} or tetracrystal SQUIDs\cite{Schulz00, Chesca02, Chesca03}, are widely recognized to provide a clear evidence for the pairing symmetry of the order parameter\cite{Tsuei00}. Such experiments rely on Josephson junctions, which for IL cuprate thin films became available only very recently\cite{Tomaschko11a}.

%
\begin{figure}[tb!] 
\begin{center}
\includegraphics[width=\columnwidth,clip]{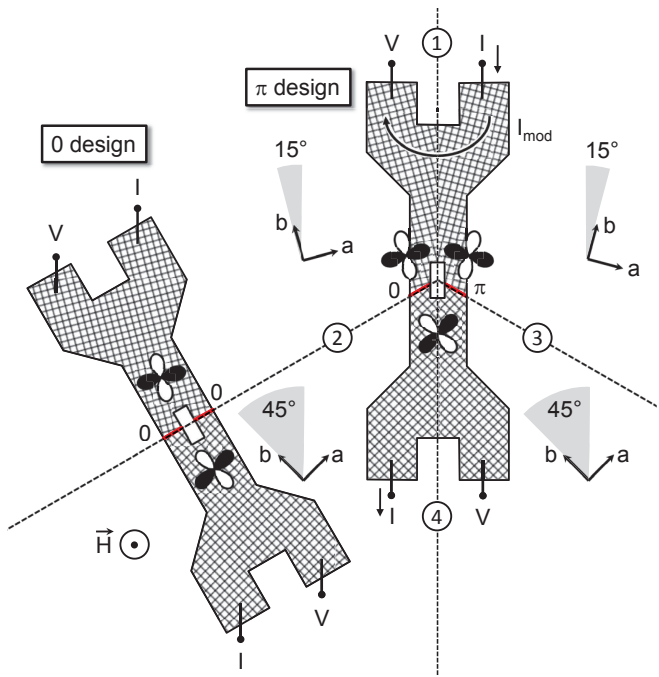}
\end{center}
\caption{(Color online)
Schematic layout of the SQUIDs. The 0-design SQUID comprises two conventional GBJs (0 junctions) straddling a single $30^\circ$ [001]-tilt grain boundary. The $\pi$-design SQUID comprises four GBs. The misorientation angle of GB 4 is $0^\circ$. All other misorientation angles are $30^\circ$. The \dxy-wave order parameter is indicated by the cloverleaf structure consisting of white and black lobes, indicating the sign change of the order parameter.
Leads for bias current $I$ and voltage $V$ are indicated. In some experiments, for the $\pi$-design SQUID we have also sent a current $I_{\rm{mod}}$ across GB 1. Magnetic fields
have been applied perpendicular to the substrate plane. 
}
\label{fig:Layout}
\end{figure}

%
%
%
 
Here we report on the fabrication and characterization of thin-film Sr$_{1-x}$La$_x$CuO$_2$ (SLCO) dc SQUIDs based on BaTiO$_3$-buffered tetracrystal SrTiO$_3$ substrates.
The geometry involved is designed to be frustrated for \dxy-wave pairing, i.e.~the SQUID ring comprising the tetracrystal point contains one 0 junction and one $\pi$ junction, if the order parameter has \dxy-wave symmetry. 
This device will be referred to as $\pi$-design SQUID. Its geometry, together with the design of a reference SQUID, is sketched in Fig.~\ref{fig:Layout}. There are four grain boundaries (GBs), labelled 1--4.  GBs 1--3 have misorientation angles of $30^\circ$, while the misorientation angle of GB 4 is $0^\circ$. The $\pi$-design SQUID comprises all GBs. GB 4 will not form a grain boundary junction (GBJ) due to its $0^\circ$ misalignment angle, in contrast to GBs 1--3. 
GBs 2 and 3, having a width of $58\,\mu$m, form the active Josephson junctions in the current and voltage lead configuration indicated in Fig.~\ref{fig:Layout}. The bias current also passes GBJ 1 which, however, is much longer ($\sim$1.5\,mm) than GBJs 2 and 3 and thus will have a much higher critical current. Below, we will see however, that flux quanta (Josephson fluxons) can enter this GBJ, which thus cannot be ignored in the data analysis.  If  SLCO is a \dxy-wave superconductor, one of GBJs 2 and 3 faces a sign change of the order parameter (GBJ 3 in Fig.~\ref{fig:Layout}), thus forming a $\pi$ Josephson junction. The other GBJs are conventional. 
The area of GBJs 2 and 3 is not much smaller than the area of the SQUID hole. In this ``spatially distributed junction'' design \cite{Chesca99} the junction's $I_c$ vs. $H$ modulation (Fraunhofer pattern) is superposed on the SQUID modulation on a similar field scale. The (a)symmetry of the SQUID modulation relative to the Fraunhofer envelope allows to detect residual fields and often also trapped magnetic flux. 
 
The reference SQUIDs -- there were two reference SQUIDs, producing very similar results -- cross only one of the $30^\circ$ GBs and incorporate two $50\,\mu$m wide GBJs, which act as conventional junctions both for s-wave and \dxy-wave order parameters. Below, these devices will be referred to as the 0-design SQUIDs.
Both the $\pi$-design SQUID and the reference SQUIDs had rectangular SQUID holes with an area $A_S = 50 \times 75\,\mu$m$^2$.

The samples have been fabricated by pulsed laser deposition, as described elsewhere\cite{Tomaschko11, Tomaschko11a, Tomaschko12}.
In brief, we first deposited a 25\,nm thick BaTiO$_3$ thin-film on the SrTiO$_3$ tetracystal, acting as a buffer layer. 
This layer was followed by a 22\,nm thick SLCO thin-film, with doping $x \approx 0.15$. Finally, a 10\,nm thick gold layer was evaporated in-situ, protecting SLCO from degradation and acting as resistive shunt for the GBJs.
The SQUIDs were patterned by standard photolithography and argon ion milling. The SLCO film had a critical temperature $T_c \approx$ 18\,K. 
Electric transport measurements were performed at $T$ = 4.2\,K in a four-point configuration, with the sample mounted inside a noise-filtered, magnetically and radio frequency shielded probe in a liquid-helium dewar.
A SQUID amplifier was used to allow for low-noise measurements.

Below we discuss data of one of the 0-design SQUIDs and of the $\pi$-design SQUID. 
The current voltage ($IV$) characteristics of these devices were nonhysteretic and could be well reproduced by the SQUID Langevin equations, extended by taking the nonzero junction width into account. Details can be found in Ref. \onlinecite{Supplement}, giving reasonable values for the junction parameters $I_0$ (maximum amplitude of Josephson current), $R$ (junction resistance) and $C$ (junction capacitance). The junctions of each device were symmetric in terms of $I_0$, $R$ and $C$, with values  $I_0$ = 8.2\,$\mu$A, $R$ = 0.92\,$\Omega$, $C$ = 24\,pF (0-design SQUID) and $I_0$ = 12.2\,$\mu$A, $R$ = 0.87\,$\Omega$, $C$ = 25\,pF ($\pi$-design SQUID).
For the inductance parameter $\beta_L = 2I_0L/\Phi_0$, where $\Phi_0$ is the magnetic flux quantum and $L$ is the total inductance of the SQUID, we found $\beta_L$ = 1.4 (0-design SQUID; $L$ = 177\,pH) and 2.2 ($\pi$-design SQUID; $L$ = 187\,pH), with an asymmetry $a_L$ = 0.05 between the left and right arm of the SQUID (both designs). A fraction $f_J$ = 0.128 (0-design SQUID) and $f_J$ = 0.12 ($\pi$-design SQUID) of the flux $\Phi$ applied to the SQUID loop was coupled to each junction. The numbers for $f_J$ were derived from an analysis of the SQUID critical current $I_c$ vs. applied field $H$ (see below).

\begin{figure}[tb!] 
\begin{center}
\includegraphics[width=0.8\columnwidth,clip]{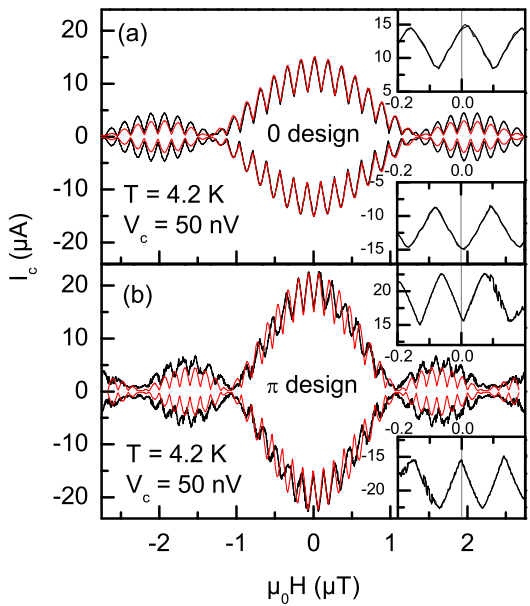}
\end{center}
\caption{(Color online) 
Critical current $I_c$ vs. applied magnetic field $H$ of (a) the 0-design SQUID and (b) the $\pi$-design SQUID (black lines) together with calculated curves (red lines) for (a) a  0 SQUID and (b) a $\pi$ SQUID. Insets show  $I_c$ vs. $H$ on expanded scales. A 50\,nV voltage criterion was used to determine $I_c$.
}
\label{fig:IcH}
\end{figure}

Fig.~\ref{fig:IcH} (a) shows the measured $I_c$ vs. $H$ for the 0-design SQUID (black line). $I_c$ was determined using a voltage criterion $V_c$ = 50\,nV. 
To identify magnetic hysteresis effects, $I_c$ vs. $H$ was traced from  2.8\,$\mu$T to -2.8\,$\mu$T and back to  2.8\,$\mu$T.
One observes a SQUID modulation with period $\mu_0\Delta H$ = 0.175\,$\mu$T on top of a Fraunhofer-like modulation, which is due to the finite junction size. No hysteresis is visible.
The modulation period corresponds to an effective SQUID area of $1.18\cdot 10^4\,\mu$m$^2$, pointing to a flux compression of about 3.15. This is reasonable for our structures\cite{Ketchen85, Chesca03}.
The insets of Fig.~\ref{fig:IcH} (a) show $I_c$ vs. $H$ near $H$ = 0 for both positive and negative $I_c$. The $I_c$ maximum is close to $H$ = 0, with a small offset of 10.6\,nT for positive $I_c$ and 5.2\,nT for negative $I_c$. We contribute the asymmetry of $\pm$2.7\,nT in offsets to an inductance asymmetry ($a_L$ = 0.05) of the two SQUID arms and the average part of 7.9\,nT to residual fields in the cryostat. The red line in Fig.~\ref{fig:IcH} (a) is a numerical calculation. It produces data very well inside the main maximum of the Fraunhofer envelope. The first Fraunhofer side-maximum is lower in amplitude than the experimental data, presumably due to a field  distribution inside the junction, which is more complex than the homogeneous flux density assumed in our model. Most importantly, however, we see that the 0-design SQUID behaves as it should be expected from a conventional 0-SQUID.

The measured $I_c$ vs. $H$  of the $\pi$-design SQUID is shown by the black line in Fig.~\ref{fig:IcH} (b). Also here we have varied $H$ from 2.8\,$\mu$T to -2.8\,$\mu$T and back to  
2.8\,$\mu$T. $I_c$ is at a \textit{minimum} near $H$ = 0 - a feature which appears when one of the two Josephson junctions \textit{exhibits an additional $\pi$ shift in its phase}. At negative $I_c$ the minimum is at $\mu_0H \approx$ 0.15\,nT, while at positive $I_c$ it appears at $\mu_0H \approx$ 7.8\,nT, pointing to an offset field of about 4\,nT and a  small asymmetry in inductance ($a_L$ = 0.05). The SQUID modulation period is $\mu_0\Delta H$ = 0.136\,$\mu$T, corresponding to an effective area of $1.52\cdot10^4\,\mu$m$^2$ and a flux compression factor of 4.05.
The overall modulation of $I_c$ vs. $H$ is described reasonably well by numerical calculations (red line), however less well than $I_c$ vs. $H$ of the 0-design SQUID. 

A prominent feature are the jumps in $I_c$, visible at $\mu_0H > $ 89\,nT at positive $I_c$ and at $\mu_0H < $ -80\,nT at negative $I_c$. There is only a very tiny hysteresis associated with these jumps, which is not even visible in Fig.~\ref{fig:IcH} (b). By comparing measured and calculated $I_c$ vs. $H$ curves within the main Fraunhofer lobe one sees that the calculated curve exhibits one additional SQUID period.
These features indicate that magnetic flux quanta enter the device at each jump. This effect was not visible for the 0-design SQUID. A strong candidate for flux entry is thus GBJ 1 which is absent in the reference SQUID. Note that the $I_c$ jumps visible in Fig.~\ref{fig:IcH} (b) occur point-symmetric, \textit{i.e.}, at positive $I_c$ they occur at positive fields, while at negative $I_c$ the field is negative, with about the same amplitude as for positive $I_c$. This feature can also clearly be seen in $V$ vs. $\Phi$ patterns taken at many values of bias current, see Ref. \onlinecite{Supplement}. 
There we also show, that the point symmetry in $I_c$ vs. $H$ holds even for large values of $H$, and that applying an additional current $I_{\rm{mod}}$ across GBJ 1 alters the values of $H$ in a way that is compatible with the notion of Josephson fluxons having entered GBJ 1.  

 
\begin{figure}[tb!] 
\begin{center}
\includegraphics[width=0.8\columnwidth,clip]{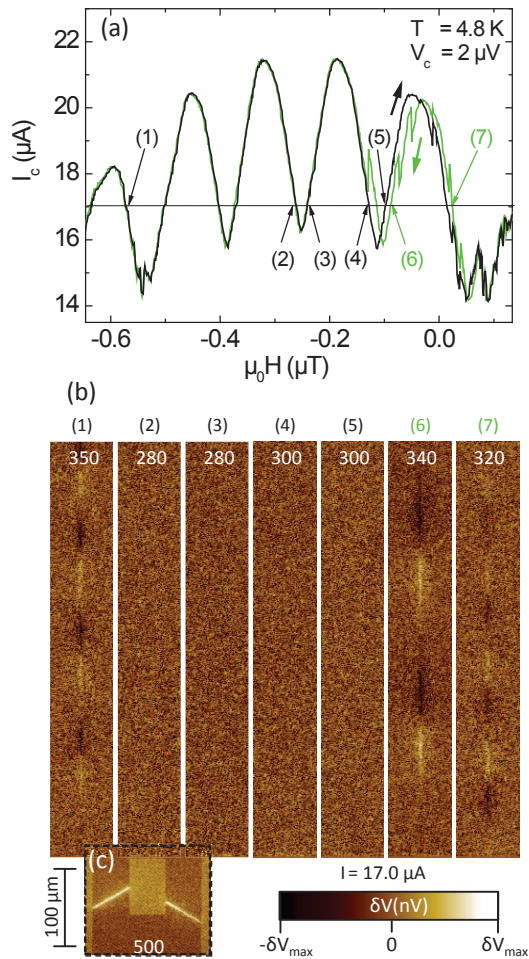}
\end{center}
\caption{(Color online) 
LTSEM data: (a) $I_c$ vs. $H$, as measured in the LTSEM setup (black and green lines distinguish sweep directions), (b) $\delta V$ images, taken along GBJ 1 at $I$ = 17\,$\mu$A [horizontal line in (a)] at the magnetic field values (1) -- (7) indicated in (a). Graph (c) shows a $\delta V$ image of the SQUID hole and GBJs 2 and 3 at $I$ = 22\,$\mu$A and $\mu_0 H$ = -0.19\,$\mu$T. This image has been superposed to scale to image (b)-(2) in order to indicate the position of the SQUID hole.
For each image $\delta V_{\rm{max}}$ is indicated inside the graphs.  
}
\label{fig:LTSEM}
\end{figure}

For a final proof we have imaged the current distribution of the $\pi$-design SQUID using low temperature scanning electron microscopy (LTSEM). Details of the method can be found in Refs. \onlinecite{Guerlich09, Straub01}. In brief, the pulsed electron beam, which is scanned across the sample, causes local heating by $\sim$ 1\,K. The measured integral quantity is the voltage $V$ across the SQUID, which is biased slightly above $I_c$. The electron beam causes a change $\delta V(x,y)$ depending on the beam position $(x,y)$.
When the beam is scanned across GBJs 2 and 3 near a $I_c$ maximum, a positive signal appears, because $I_c$ is lowered, causing a slight increase of $V$ (cf. Fig.~\ref{fig:LTSEM} (c)). 
When GBJ 1 is free of vortices no signal is expected from this GB. By contrast, when Josephson fluxons are present, local heating will alter the screening currents around the fluxons. In a heated area the Cooper pair density and thus the maximum supercurrent density is suppressed, causing an increase of the Josephson length $\lambda_J$ and, thus, the fluxon is virtually deformed towards the heated area. When the electron beam is between the fluxon center and the SQUID hole this causes an increase of the fluxon's stray flux coupled to the SQUID and thus a change in $I_c$. In the opposite case the stray flux is decreased. A fluxon will thus appear as a bipolar signal $\delta V(x,y)$, with increased/decreased voltage relative to the unperturbed value of $V$. In a similar way, Abrikosov vortices trapped in a YBa$_2$Cu$_3$O$_7$ SQUID\cite{Straub01}, as well as Josephson fluxons having entered a GBJ\cite{Koelle00}, have been imaged by LTSEM.

Fig.~\ref{fig:LTSEM} (a) shows $I_c$ vs. $H$, as measured in the LTSEM setup at $T$ = 4.8\,K using a voltage criterion $V_c$ = 2\,$\mu$V. There is a stronger offset field ($\sim 0.25\,\mu$T) than in the transport setup, causing a shift by about 2 SQUID modulation periods. Due to the Fraunhofer envelope this shift is straightforward to recognize. The two sweep directions of $H$ have been distinguished by black and green lines. There is a nonhysteretic region around the offset field; at larger values of $H$, $I_c$ jumps occur, leading to magnetic hysteresis. 
$\delta V$ images of GBJ 1, cf. Fig.~\ref{fig:LTSEM} (b), have been taken at $I$ = 17\,$\mu A$ at the field values indicated by labels (1)--(7) in  Fig.~\ref{fig:LTSEM} (a). 
No contrast appears when the sample is biased near the offset field [images (2), (3)] or at a field smoothly extending this $I_c$ vs. $H$ pattern [images (4),(5)]. By contrast, when $I_c$ jumps have occurred, we observed a periodically modulated signal, having a period decreasing with the field amplitude $H$ relative to the offset field [images (1),(6),(7)]. This is very indicative of Josephson fluxons having entered GBJ 1. 
In Fig.~\ref{fig:LTSEM} (c) we also show a $\delta V$ image of the area around the SQUID hole and GBJs 2 and 3. The image has been taken in a separate run, because we did not want to disturb the images of  Fig.~\ref{fig:LTSEM} (b)  by scanning across these GBJs. It has been taken at the maximum of the SQUID modulation at $I$ = 22\,$\mu$A and $\mu_0 H$ = -0.19\,$\mu$T. The image is superposed to scale to image Fig.~\ref{fig:LTSEM} (b)-(2), to give an impression of the position of the SQUID hole and GBJs 2 and 3 relative to the images of Fig.~\ref{fig:LTSEM} (b).

The LTSEM data clearly show that the $I_c$ vs. $H$ region of interest is free of trapped flux; we thus feel safe in interpreting the $\pi$-design data in favor of a \dxy-wave symmetry of  Sr$_{1-x}$La$_{x}$CuO$_{2}$. 

One may in addition ask about subdominant order parameters. A real superposition \dxy$\pm s$ is not very likely due to the tetragonal crystal symmetry but, if present, could lead to an asymmetry of the critical currents of GBJs 2 and 3 and, in consequence, to a similar shift as the one we interpreted in terms of an inductance asymmetry. On the other hand, the 0-design SQUID should not show this asymmetry and we thus believe that an inductance asymmetry is more likely. 
By contrast, a complex admixture of a subdominant order parameter would lead to a ground state phase different from $\pi$ or 0. Then, the SQUID modulation would shift relatively to the Fraunhofer envelope, making the amplitude 
of the inner $I_c$ maxima asymmetric. This effect is not observed at least on a $\sim 5\%$ level.


In summary, our data clearly show, that the superconducting order parameter of the electron doped infinite-layer high-$T_c$ cuprate Sr$_{1-x}$La$_{x}$CuO$_{2}$ has \dxy-wave symmetry. The phase sensitive configuration used was a $\pi$ SQUID patterned on a tetracrystal. The parasitic effect of Josephson fluxons entering one of the grain boundary junctions has been ruled out by direct imaging of the local supercurrent contribution.
Sr$_{1-x}$La$_{x}$CuO$_{2}$ has the most simple crystal structure of all high-$T_c$ cuprates. We conclude that the \dxy-wave symmetry is inherent to cuprate superconductivity and neither restricted to hole doping nor related to the complex crystal structures that complicates an analysis of almost all other cuprate superconductors.

\acknowledgments
{
J.~T. gratefully acknowledges support by the Evangelisches Studienwerk e.V. Villigst and  J.~N.  by the Carl-Zeiss Stiftung.
V.~L. acknowledges partial financial support by a grant of the Romanian National Authority for Scientific Research, CNCS UEFISCDI, project number PN-II-ID-PCE-2011-3-1065.
This work was funded by the Deutsche Forschungsgemeinschaft (project KL 930/11).
}

\bibliography{SLCO_bibliography}

\end{document}



\title{Supplementary information to \\``Phase-sensitive evidence for $d_{x^2-y^2}$ - pairing symmetry in the\\ parent-structure high-$T_c$ cuprate superconductor Sr$_{1-x}$La$_{x}$CuO$_{2}$''}

\author{J.~Tomaschko}
\affiliation{%
  Physikalisches Institut -- Experimentalphysik II and Center for Collective Quantum Phenomena in LISA$^+$,
  Universit\"{a}t T\"{u}bingen,
  Auf der Morgenstelle 14,
  72076 T\"{u}bingen, Germany
}
\author{S.~Scharinger}
\affiliation{%
  Physikalisches Institut -- Experimentalphysik II and Center for Collective Quantum Phenomena in LISA$^+$,
  Universit\"{a}t T\"{u}bingen,
  Auf der Morgenstelle 14,
  72076 T\"{u}bingen, Germany
}
\author{V.~Leca}
\affiliation{%
  Physikalisches Institut -- Experimentalphysik II and Center for Collective Quantum Phenomena in LISA$^+$,
  Universit\"{a}t T\"{u}bingen,
  Auf der Morgenstelle 14,
  72076 T\"{u}bingen, Germany
}
\affiliation{%
  National Institute for Research and Development in Microtechnologies, 
  Molecular Nanotechnology Laboratory, 
  Erou Iancu Nicolae Str. 126A, RO-077190, Bucharest, Romania
}
\author{J.~Nagel}
\affiliation{%
  Physikalisches Institut -- Experimentalphysik II and Center for Collective Quantum Phenomena in LISA$^+$,
  Universit\"{a}t T\"{u}bingen,
  Auf der Morgenstelle 14,
  72076 T\"{u}bingen, Germany
}
\author{M.~Kemmler}
\affiliation{%
  Physikalisches Institut -- Experimentalphysik II and Center for Collective Quantum Phenomena in LISA$^+$,
  Universit\"{a}t T\"{u}bingen,
  Auf der Morgenstelle 14,
  72076 T\"{u}bingen, Germany
}
\author{T.~Selistrovski}
\affiliation{%
  Physikalisches Institut -- Experimentalphysik II and Center for Collective Quantum Phenomena in LISA$^+$,
  Universit\"{a}t T\"{u}bingen,
  Auf der Morgenstelle 14,
  72076 T\"{u}bingen, Germany
}

\author{D.~Koelle}
\author{R.~Kleiner}
\email{kleiner@uni-tuebingen.de}
\affiliation{%
  Physikalisches Institut -- Experimentalphysik II and Center for Collective Quantum Phenomena in LISA$^+$,
  Universit\"{a}t T\"{u}bingen,
  Auf der Morgenstelle 14,
  72076 T\"{u}bingen, Germany
}
\date{\today}

\maketitle

In this supplement we provide additional experimental data for the 0-design SQUID and the $\pi$-design SQUID: (I) current voltage ($IV$) characteristics, (II) voltage $V$ vs. applied field $H$, (III) symmetry considerations on $I_c$ vs. $H$, and (IV) a measurement of the current step height caused by the SQUID $LC$ resonances. Most measurements are accompanied by simulations. The model equations are described in section I.

\section{Current voltage characteristics and model}
\label{sec:IV}

\begin{figure}[tb!] 
\begin{center}
\includegraphics[width=0.9\columnwidth,clip]{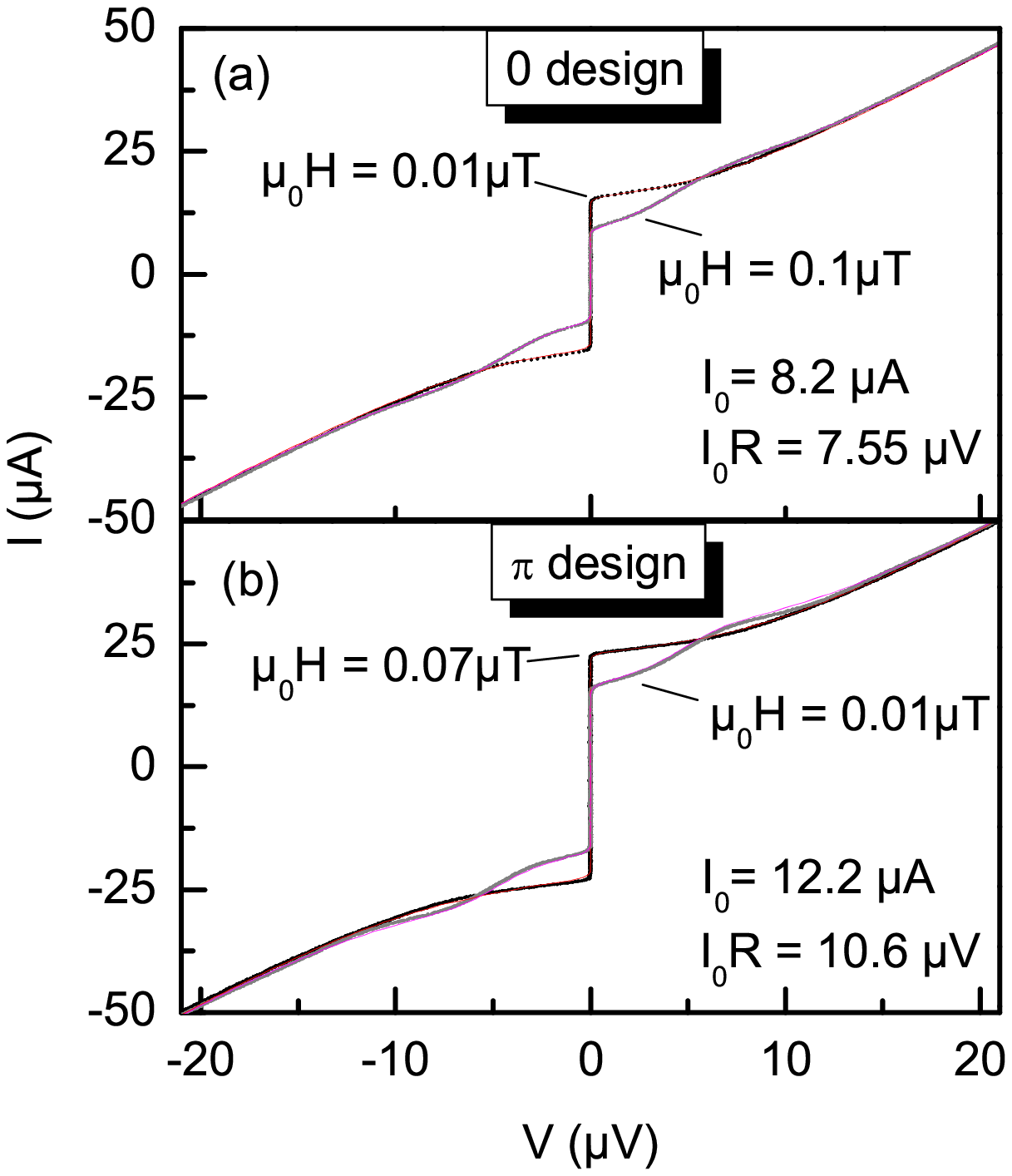}
\end{center}
\caption{(Color online)
Current voltage ($IV$) characteristics of (a) the 0-design SQUID  and (b) the $\pi$-design SQUID. Each graph contains two measured curves and two calculated curves.
Measurements were taken at, respectively, maximum (black circles) and minimum (grey circles) critical current. Calculated curves were obtained by numerically solving the SQUID Langevin equations, extended by taking the nonzero junction width into account, for (a) a 0 SQUID at flux $\Phi = 0$ (red line) and $\Phi = \Phi_0/2$ (magenta line), and (b) for a $\pi$ SQUID at flux $\Phi = \Phi_0/2$ (red line) and $\Phi = \Phi_0$ (magenta line). Model parameters in (a) are $\beta_c$ = 0.5, $\beta_L$ = 1.4, $\Gamma$ = 0.022, $a_L = 0.05$ and $f_J$ = 0.128. By matching abscissa and ordinate of the measured and calculated curves one obtains $I_0$ = 8.2 $\mu$A and $I_0R$ = 7.55\,$\mu$V. Model parameters in (b) are $\beta_c$ = 0.7, $\beta_L$ = 2.2, $\Gamma$ = 0.0147, $a_L$ = 0.05, $f_J$ = 0.12, $I_0$ = 12.2\,$\mu$A and $I_0R$ = 10.6\,$\mu$V. The same set of parameters was used for all curves discussed in this paper.
}
\label{fig:IV}
\end{figure}

We start to characterize our devices by discussing their current voltage ($IV$) characteristics.
Fig.~\ref{fig:IV} (a) shows two $IV$ characteristics of the 0-design SQUID. The magnetic field $H$, applied perpendicularly to the substrate plane, was adjusted such that the SQUID's critical current was at its first (counted from zero applied field) maximum (black circles, $\mu_0H$ = 0.01\,$\mu$T) or at its first minimum (grey circles, $\mu_0H$ = 0.1\,$\mu$T). In both cases the $IV$ characteristics were nonhysteretic. They can be very well fitted by numerically calculated curves, solving the SQUID Langevin equations\cite{Clarke04}, extended by taking the nonzero junction width into account. The model assumes that the Josephson junctions can be described by the resistively and capacitively shunted junction model\cite{Stewart68, McCumber68}. We have assumed further, that a fraction $f_J$ of the flux $\Phi$ through the SQUID loop is homogeneously coupled into the junctions, causing a linear increase of the Josephson phase differences $\gamma_k$ ($k$ = 1,2) inside the junctions. This contribution has been integrated out analytically, yielding a sinc function for the critical current $I_0$ vs. $\Phi$ for each junction. 

The normalized currents $i = I/I_0$ through the junctions are given by

\begin{equation}
\frac{i}{2} \pm j = \beta_c \ddot{\gamma_k} + \dot{\gamma_k} + i_{c}(\Phi)\sin(\gamma_k)+i_{N,k}
\label{Eq:RSJ1}
\end{equation}
where $k$ = 1,2 labels the two Josephson junctions. $I_0$ is the amplitude of the Josephson current, $j = J/I_0$ is the normalized circulating current in the SQUID loop and  `$\pm$' refers to junctions 1 and 2, respectively. 
$\beta_c = 2\pi I_0R^2C/\Phi_0$ is the Stewart-McCumber parameter. $\Phi_0$ is the flux quantum and $R$ and $C$, respectively, denote junction resistance and capacitance.
$\gamma_k$ is the Josephson phase difference across junction $k$ and the dots denote derivative with respect to time. The flux dependent quantity $i_{c}(\Phi)$ 
is given by $i_{c}(\Phi)$ = sin$(\pi f_J\Phi/\Phi_0)/[\pi f_J\Phi/\Phi_0]$. 
The normalized noise current $i_{N,k}$ has a spectral power density $4\Gamma$, with $\Gamma = 2\pi k_BT/I_0\Phi_0$. The above equations assumed that the junction parameters $\beta_c$ and $i_{c}(\Phi)$ are the same for both junctions. 
If junction $k$ is a $\pi$ junction, a phase $\pi$ is to be added to $\gamma_k$.

The two phases $\gamma_k$ are related by

\begin{equation}
\gamma_2 - \gamma_1 = 2\pi\Phi/\Phi_0 + \pi\beta_L (j + a_Li)
\label{Eq:RSJ2}
\end{equation}
where $\beta_L = 2I_0L/\Phi_0$. $L = L_1+L_2$, where $L_1$ and $L_2$ are the inductances of the two SQUID arms, related to the inductance asymmetry $a_L$ via $L_k = L(1\pm a_L)/2$.  

From Eqs.~\eqref{Eq:RSJ1} and \eqref{Eq:RSJ2} one obtains the normalized voltage $v = V/I_0R$, and thus current voltage characteristics, critical current vs. flux etc., by taking the time average of $(\dot{\gamma_1}+\dot{\gamma_2})/2$.
A consistent set of model parameters can be obtained by analyzing $IV$ characteristics at maximum and minimum $I_c$, plus $I_c$ vs. $H$. 

The calculated $IV$ characteristics in Fig.~\ref{fig:IV} (a) are for $\beta_c$ = 0.5, $\beta_L$ = 1.4, $\Gamma$ = 0.022, $a_L = 0.05$ and $f_J$ = 0.128 (the latter two numbers are actually determined from $I_c$ vs. $H$ data).
In dimensioned units one finds $I_0$ = 8.2\,$\mu$A and $I_0R$ = 7.55\,$\mu$V, $R$ = 0.92\,$\Omega$, $L$ = 177\,pH, $C$ = 24\,pF. These are reasonable numbers for our SQUIDs.
In the $\mu_0H$ = 0.1\,$\mu$T curve of Fig.~\ref{fig:IV} (a) one also notices a hump for 6\,$\mu$V  $< V <$ 10\,$\mu$V. This is a $LC$ resonance, which becomes maximally excited when the supercurrents across the two junctions oscillate out-of-phase\cite{Clarke04}. 

Two $IV$ characteristics for the $\pi$-design SQUID  are shown in Fig.~\ref{fig:IV} (b). One first notices that the first $I_c$ maximum (black circles) was obtained at a relatively large field, 0.07\,$\mu$T. By contrast, at $\mu_0H$ = 0.01\,$\mu$T, $I_c$ had a minimum (grey circles). Simulating these curves (red line for $\mu_0H$ = 0.07\,$\mu$T and magenta line for $\mu_0H$ = 0.01 $\mu$T) we have assumed that junction 2 is a $\pi$ junction and further used the parameters $\beta_c$ = 0.7, $\beta_L$ = 2.2, $\Gamma$ = 0.0147, $a_L$ = 0.05, $f_J$ = 0.12, $I_0$ = 12.2\,$\mu$A, $I_0R$ = 10.6\,$\mu$V, $R$ = 0.87\,$\Omega$, $C$ = 25\,pF, $L$ = 187\,pH, which are not very different from the reference SQUID.

\section{Voltage vs. Applied Field}
\label{sec:VH}

\begin{figure}[tb!] 
\begin{center}
\includegraphics[width=\columnwidth,clip]{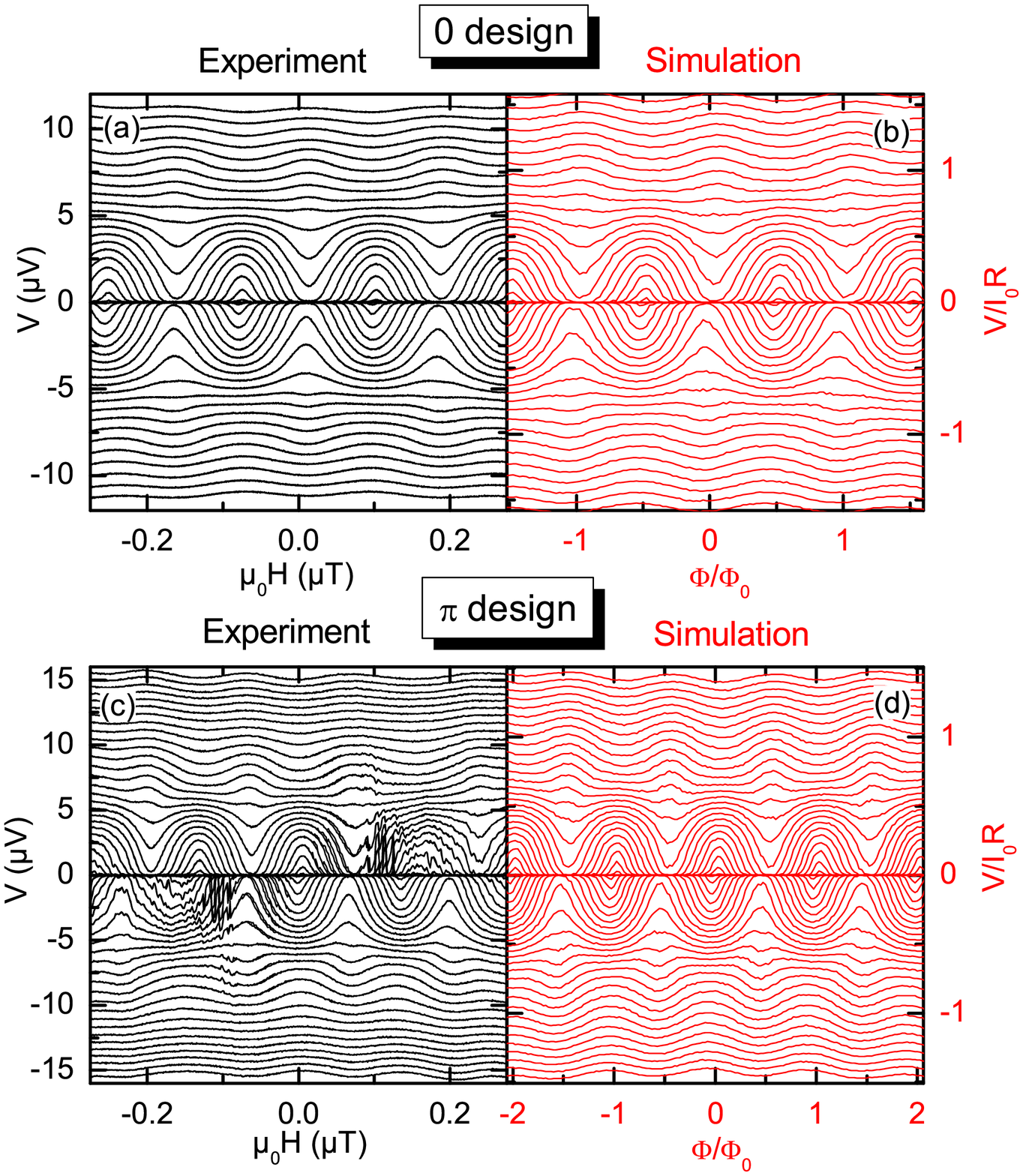}
\end{center}
\caption{(Color online) 
Measured $V$ vs. $H$ for (a) the 0-design SQUID and (c) the $\pi$-design SQUID together with numerically calculated curves for (c) the 0-design SQUID and (d) the $\pi$-design SQUID.
The current $I$ across the 0-design SQUID has been increased (decreased) from 0 in steps of 1\,$\mu$A up to $\pm$28\,$\mu$A. For the $\pi$-design SQUID the step width was 0.98\,$\mu$A, with a maximum current $\pm$39.5\,$\mu$A. 
}
\label{fig:V_Phi}
\end{figure}

Fig.~\ref{fig:V_Phi} shows measured [(a), (c)] and calculated [(b),(d)] patterns $V$ vs. $H$ for the 0-design SQUID [(a), (b)] and the $\pi$-design SQUID [(c),(d)].
The data were obtained in the same run as the $IV$ characteristics of Fig.~\ref{fig:IV} and the $I_c$ vs. $H$ data (Fig. 2 of the main paper).
In the measurements the current $I$ across the 0-design SQUID has been increased (decreased) from 0 in steps of 1 $\mu$A up to $\pm$28 $\mu$A. For the $\pi$-design SQUID the step width was 0.98\,$\mu$A, with a maximum current $\pm$39.5\,$\mu$A. Measured data for the 0-design SQUID, cf. Fig.~\ref{fig:V_Phi} (a),  are smooth and well reproduced by the calculated patterns [Fig.~\ref{fig:V_Phi} (b)]. Note that for voltages below $|V| \approx$ 5 $\mu$V the $|V|$ vs. $H$ minima are located near $H$ = 0 (modulo modulation period), while for $|V| >$ 5\,$\mu$V one finds a maximum here. This is due to the $LC$ resonance which is also visible in Fig.~\ref{fig:IV} (a). Further, the $|V|$ vs. $H$ maxima appear (modulo modulation period) near $\mu_0H$ = 0.995\,$\mu$T and are slightly shifted with respect to the origin of the ordinate. This shift, which is due to the small inductance asymmetry, was seen already in $I_c$ vs. $H$. Otherwise, the device behaves as it can be expected for a 0 SQUID. 
In the experimental data for the the $\pi$-design SQUID one observes similar features, but shifted by a half period with respect to the $\pi$-design SQUID. As in $I_c$ vs. $H$, jumps appear in the second SQUID modulation period at positive $H$ at positive $V$ and at negative $H$ at negative $V$. This point symmetry was already visible in the $I_c$ vs. $H$ data. Apart from these jumps, $V$ vs. $H$ is reasonably well reproduced by simulations, cf. Fig.~\ref{fig:V_Phi} (b).

\section{Critical Current vs. Applied Field: Symmetry Considerations}
\label{sec:Ic_H_Symmetry}

\begin{figure}[tb!] 
\begin{center}
\includegraphics[width=\columnwidth,clip]{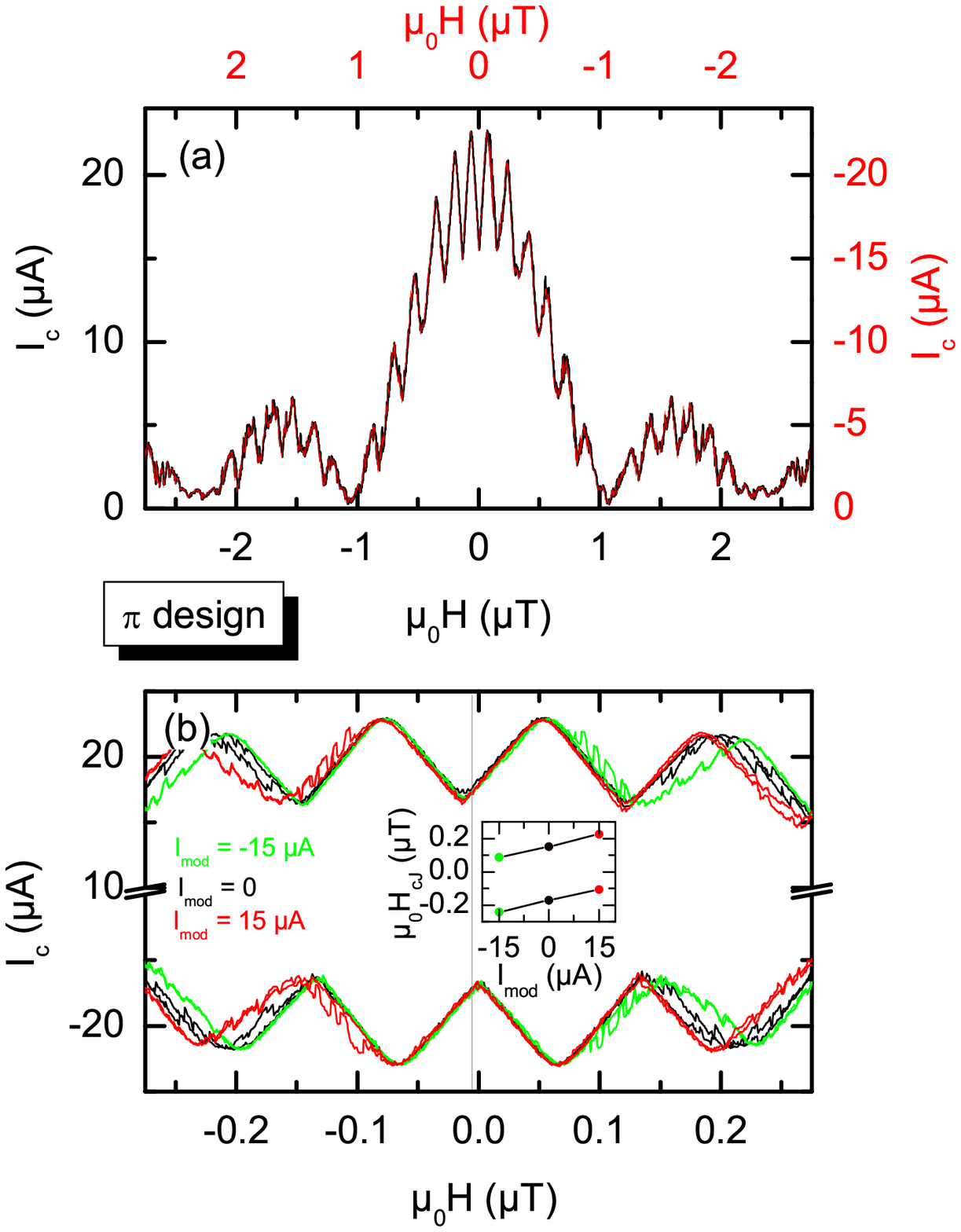}
\end{center}
\caption{(Color online)
$I_c$ vs. $H$ of the $\pi$-design SQUID: (a) Test of symmetries of the $I_c$ jumps. Plot of negative $I_c$ as $-I_c$ vs. $-\mu_0H$ (red line) on top of $I_c$ vs. $\mu_0H$ for positive $I_c$ (black line), demonstrating point symmetry. (b) $I_c$ vs. $H$  with an additional modulation current $I_{\rm{mod}}$ applied across GBJ 1. Black line: $I_{\rm{mod}} = 0$; green line: $I_{\rm{mod}} = -15\,\mu$A; red line: $I_{\rm{mod}} = 15\,\mu$A. Inset shows first penetration fields $\mu_0H_{\rm{cJ}}$ vs. $I_{\rm{mod}}$.  Grey line in (b) indicates offset field.
}
\label{fig:IcH_Symmetries}
\end{figure}

In Fig.~\ref{fig:IcH_Symmetries} (a) we further demonstrate point symmetry by  plotting the negative $I_c$ as $-I_c$ vs. $-\mu_0H$ (red line) on top of $I_c$ vs. $\mu_0H$ for the positive $I_c$ (black line). The curves are basically indistinguishable.

The point symmetry gives some hint on the properties of the trapped flux. First, if Abrikosov vortices or Josephson fluxons were trapped permanently one would at most expect a symmetry with respect to a change $I \leftrightarrow -I$. The point symmetry implies that also the \textit{polarity} of the trapped flux changes by reversing the magnetic field.
The fact that the $I_c$ jumps exhibit almost no hysteresis further implies that almost no pinning is present. All this, together with the observation that jumps in $I_c$ vs. $H$ are absent for the 0-design SQUID, supports the suspicion that Josephson fluxons enter and leave GBJ 1. The change in polarity of the trapped flux with applied field implies that \textit{no fluxons }are present at low fields, allowing to interpret our data in terms of a $\pi$ SQUID and thus in favor of a \dxy-wave order parameter symmetry.

\begin{figure}[tb!] 
\begin{center}
\includegraphics[width=0.9\columnwidth,clip]{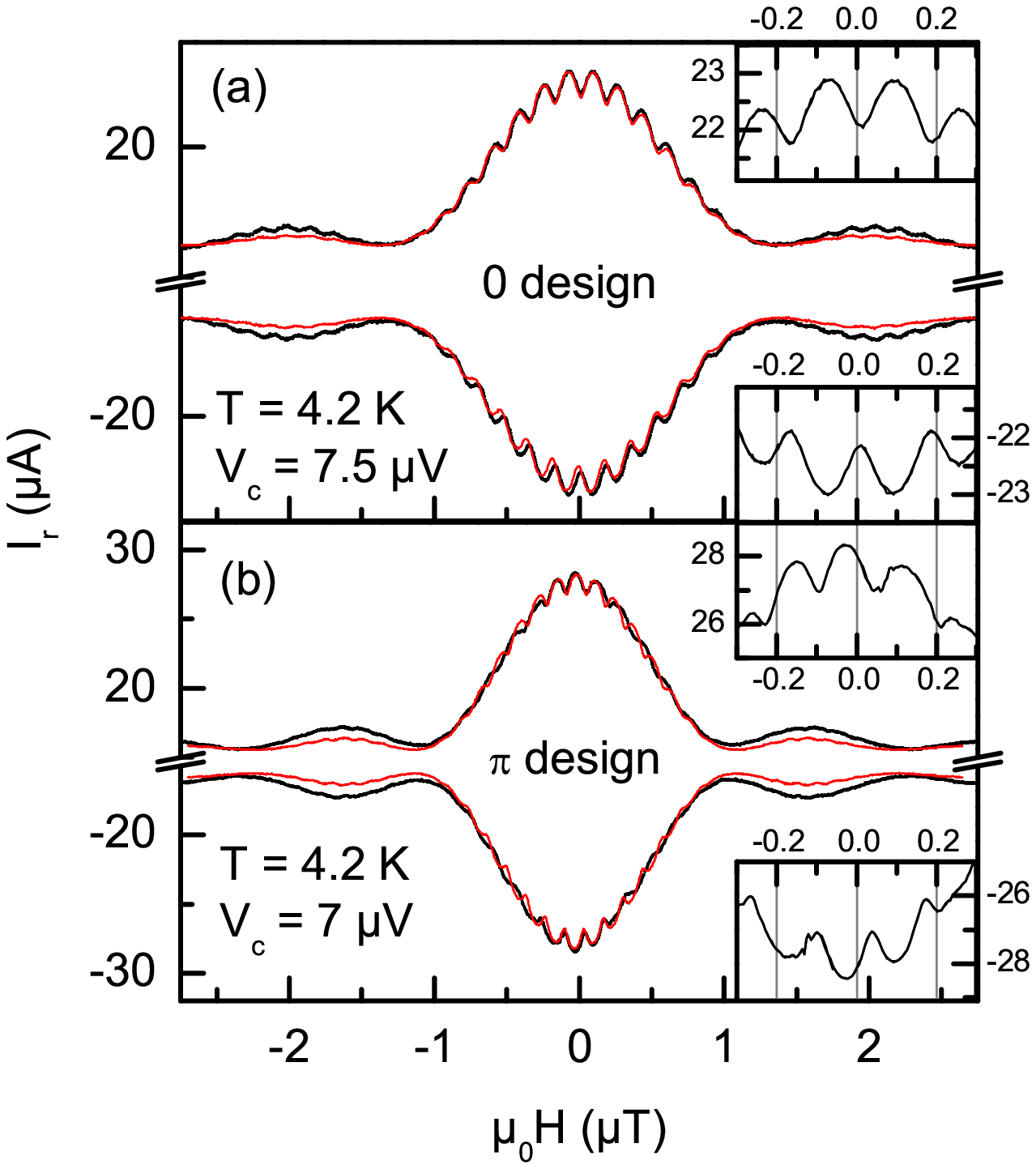}
\end{center}
\caption{(Color online)
Resonance current $I_r$ vs. applied field for (a) the 0-design SQUID and (b) the $\pi$-design SQUID (black curves), together with the corresponding simulated curves (red lines). Insets show $I_r$ vs. $H$ at expanded current and field scales.
}
\label{fig:LC_Reso}
\end{figure}


The $I_c$ jumps in Fig. 2 (b) of the main paper were strongly asymmetric with respect to the applied field. For positive current the first $I_c$ jumps occurred at the penetration fields $\mu_0H_{cJ}$ = +0.09\,$\mu$T and at -0.24\,$\mu$T. At negative current they were observable at $\mu_0H_{cJ}$ = -0.09\,$\mu$T and at +0.24\,$\mu$T.
This asymmetry was different in different cooldowns and seemed to depend on the residual field. In fact, screening currents across GBJ 1 will change the surface barrier for fluxon entry, making $H_{\rm{cJ}}$ asymmetric. Fig.~\ref{fig:IcH_Symmetries} (b) shows results from another cooldown where the jumps occurred almost symmetrically with respect to $H$ (black line). In this graph we also show two measurements, where we applied an additional modulation current $I_{\rm{mod}}$ across GBJ 1, cf. Fig. 1 in the main paper. Depending on the polarity of this current its effect was to linearly shift the appearance of the $I_c$ jumps, \textit{i.e.} $H_{\rm{cJ}}$, to higher (lower) values for positive (negative) values of $I_{\rm{mod}}$. This shift did not depend on the sign of $I$. 
$I_{\rm{mod}}$ exerts a Lorentz force on the fluxons. For $I_{\rm{mod}} > 0$ fluxons appearing at $H > 0$ are repelled from the SQUID, increasing the switching fields for both polarities of $I$. The force on antifluxons appearing at $H < 0$ points inward, \textit{i.e.}, the switching field decreases. For $I_{\rm{mod}} < 0$ the forces on fluxons and antifluxons are opposite. Thus, the shifts of the $I_c$ jumps are fully compatible with the notion of (anti)fluxons having entered GBJ 1. 

\section{\textit{LC} resonances}
\label{sec:LC}

We finally address in more detail the resonance feature which appeared as a hump in the $IV$ characteristics at $\mu_0H$ = 0.1\,$\mu$T for the 0-design SQUID and at 
$\mu_0H$ = 0.01\,$\mu$T for the $\pi$-design SQUID. In $V$ vs. $H$ the resonance appeared as a phase shift of a half period in the SQUID modulation. 
These effects are due to resonantly excited circulation currents across the SQUID loop and occur when the Josephson currents across the two junctions acquire an out-of phase component. Maximum excitation occurs when the Josephson currents oscillate maximally out-of-phase. For a 0 SQUID one thus expects the strongest effect for a half flux quantum applied to the junction, for a $\pi$ SQUID it should be strongest at integer multiples of $\Phi_0$ \cite{Chesca02}. 
To investigate the $LC$ resonance vs. applied field $H$ we have determined the ``resonance current'' $I_r$ across the SQUIDs by using a voltage criterion $V_c$ which corresponds to the center of the hump feature in the $IV$ characteristics (7.5\,$\mu$V for the 0-design SQUID and 7\,$\mu$V for the $\pi$-design SQUID). 

Fig.~\ref{fig:LC_Reso} (a) shows $I_r$ vs. $H$ for the 0-design SQUID (black line) together with a simulated curve (red line). Apart from a slight offset due to residual fields the minimum in $I_r$ is at zero field both for negative and positive values of $I_r$. By contrast, for the $\pi$-design SQUID one finds (except for an offset) a maximum of $I_r$ near zero field, again indicative of a $\pi$ SQUID.

\bibliography{SLCO_bibliography}